# Electrical Probing of Field-Driven Cascading Quantized Transitions of Skyrmion Cluster States in MnSi Nanowires

*Haifeng Du,[1] Dong Liang,[2] Chiming Jin,[1] Lingyao Kong,[3] Matthew J. Stolt,[2] Wei Ning,[1] Jiyong Yang,[1] Ying Xing,[4] Jian Wang,[4] Renchao Che,[5] Jiadong Zang,[6]\* Song Jin,[2]\* Yuheng Zhang,[1,7] Mingliang Tian[1,7]\**

[1]High Magnetic Field Laboratory, Chinese Academy of Science (CAS), Hefei, Anhui Province 230031, China

[2]Department of Chemistry, University of Wisconsin—Madison, 1101 University Avenue, Wisconsin 53706, USA

[3]Institute of Fluid Physics, China Academy of Engineering Physics, Mianyang, Sichuan Province 621900, China

[4]International Center for Quantum Materials, School of Physics, Peking University, Beijing 100871, China

[5] Advanced Materials Laboratory, Fudan University, Shanghai 200433, P. R. China

[6]Institute for Quantum Matter and Department of Physics and Astronomy, Johns Hopkins University, Baltimore, Maryland 21218, USA

[7] Collaborative Innovation Center of Advanced Microstructures, Nanjing, Jiangsu Province 210093, China

*Corresponding authors: jiadongzang@gmail.com (J.Z.); jin@chem.wisc.edu (S.J.); tianml@hmfl.ac.cn (M.T.).*

**Abstract**

Magnetic skyrmions are topologically stable whirlpool-like spin textures that offer great promise as information carriers for future ultra-dense memory and logic devices[1-4]. To enable such applications, particular attention has been focused on the skyrmions properties in highly confined geometry such as one dimensional nanowires[5-8]. Hitherto it is still experimentally unclear what happens when the width of the nanowire is comparable to that of a single skyrmion. Here we report the experimental demonstration of such scheme, where magnetic field-driven skyrmion cluster (*SC*) states with small numbers of skyrmions were demonstrated to exist on the cross-sections of ultra-narrow single-crystal MnSi nanowires (*NW*s) with diameters, *d*, (40 – 60 *nm*) comparable to the skyrmion lattice constant (18 *nm*). In contrast to the skyrmion lattice in bulk MnSi samples, the skyrmion clusters lead to anomalous magnetoresistance (*MR)* behavior measured under magnetic field parallel to the *NW* long axis, where quantized jumps in *MR* are observed and directly associated with the change of the skyrmion number in the cluster, which is supported by Monte Carlo simulations. These jumps show the key difference between the clustering and crystalline states of skyrmions, and lay a solid foundation to realize skyrmion-based memory devices that the number of skyrmions can be counted via conventional electrical measurements.

The emergence of topological phenomena and topological materials has attracted growing interests in condensed matter physics[9]. The interplay between topology and geometry not only deepens our physical understanding, but also provides new routes to device designs that incorporate novel physics. In magnetic materials, a notable example of a topologically stable object is a "skyrmion", a swirl-like spin texture that carries quantized topological charge[5]. Skyrmions have been realized in several helimagnets with the non-centrosymmetric B20 crystal structure, such as MnSi, FeGe, and $Fe_xCo_{1-x}Si$[1,4,5]. The peculiar twists of the magnetization within the skyrmion originate from the competition between the chiral Dzyaloshinskii-Moriya (*DM*) interaction and the ferromagnetic (*FM*) exchange interaction. The ratio of these two interactions determines the skyrmion size, which is typically on the order of 5-100 *nm*, and can be continuously tuned by doping[10]. Moreover, recent experiments found that skyrmions can be manipulated by electric currents with a current density several orders of magnitude lower than that needed to drive ferromagnetic domain walls[3,11]. Although these properties are advantageous for high-density data storage applications with low dissipation, most skyrmions in bulk samples or two dimensional (2D) films reported in literatures have condensed into lattices[5], which makes the single bit operation impossible. Experimental investigations of epitaxial MnSi/Si(111) films in strong in-plane magnetic field have discovered distorted skyrmion strings[12], wherein the strain-induced magnetic anisotropy by the Si substrate is crucial for the stabilization of spin textures, and skyrmions are still aligned in one dimension. While recent study enables an electric reading of the number of turns in the helical structure in MnSi thin films[13], the detection of creation and annihilation of individual skymions in devices by an electric probe that can be more readily integrated into conventional electronic architectures still lacks[14].

Single crystal nanowires (*NW*s) of the compounds with the B20 structure provides a valuable insight into this issue[15]. Skyrmion states have been identified in B20 MnSi *NW*s by Lorentz transmission electron microscopy[16] and longitudinal magnetoresistance (*MR*) measurements[17]. However, the widths of the *NW*s ($d > 200$ *nm*) in these studies are much larger than the skyrmion lattice constant ($L_d \sim 18$ *nm*) for MnSi, implying that the skymion state therein still condenses into the lattice as in bulk samples. Fortunately, our recent advances in nanomaterials allow us to synthesize high-quality crystalline MnSi *NW*s of various diameters down to tens of nanometers comparable to the skyrmion size, and conveniently manipulate the thin *NW*s (see Figures S1 in the Supplementary Information for detailed descriptions of the technology). It is thus an interesting question whether skyrmions can survive in such narrow *NW*s and behave distinctly.

In this work, we demonstrate the formation of *individual* skyrmions in thin MnSi *NW*s monitored by *MR* measurements. A skyrmion cluster[18,19], composed of sparsely distributed skyrmions rather than skyrmion lattices, is demonstrated in this confined system under an external magnetic field, *B*, aligned along the axis of the wire and confirmed by Monte Carlo simulations. The *MR(B)* curves exhibit striking jumps at specific fields that corresponds the creation or annihilation of individual skyrmions. Moreover, the number of jumps depends on the *NW* diameter, and eventually disappears both in wider and ultra-narrow *NW*s. Therefore the jumps in the *MR* curves reveal the physical signature of the cascading transition of the skyrmion cluster states, and enable an electric reading of single skrymions.

MnSi *NW*s were synthesized by chemical vapor deposition[15]. A representative scanning electron microscopy (*SEM*) image of a single wire with a smooth (111) surface is shown in Figure 1a. The cross-section of the *NW* shows a merohedral twinning structure with the (001) twin plane parallel with the <110> growth direction (Figure 1b). This is a common feature of *NW*s of non-centrosymmetric B20 compounds, in which the unique (001) twin plane partitions the *NW* into two parts with opposite handedness[20]. A high resolution transmission electron microscopy (*TEM*) image (Figure 1c), together with previous *TEM* diffraction results[15], confirms the perfect B20 crystal structure. *MR* measurements were carried out by a standard 4-probe technique on individual *NW*s (the device is illustrated in Figure S2 of the Supplementary Information). We selected a 40 *nm* diameter *NW*, the size of which is comparable to $L_d$ ~18 *nm*, to demonstrate the central results. Both resistance (*R*) versus temperature (*T*) and *MR-T* curves are shown in Figure 1d, where *MR* at field, *B*, is defined as *MR=[R(B)-R(0 Oe)]/R(0 Oe)*. The residual resistivity ratio of the wire is *R(300 K)/R(5 K)* ~ 7.5. The Curie temperature $T_c$ of the *NW* is determined to be ~ 29 *K* from Figure 1d, which is almost the same as that of bulk MnSi (29.5 *K*). The room temperature resistivity, $\rho \approx 198 \mu\Omega \cdot cm$, is quite close to the value of $180 \mu\Omega \cdot cm$ reported for bulk single-crystal MnSi[21]. These results reflect the high quality of the *NW*s.

Figure 2 shows a typical curve of *MR* as a function of *B* for the 40 *nm NW* at $T = 14$ *K*. The origin of the *MR* is ascribed to the coupling of conduction electrons with chiral modulations[22,23]. Previous theoretical calculation[24] and experimental works on MnSi samples[17,25,26] have well established that the magnetic scattering of the conduction electrons by the skyrmion phase would lead to two kinks in the *MR(B)* isotherms, which correspond to the lower and upper critical fields driving the system into and out the skyrmion phase. The data were recorded by increasing the field from -8 *kOe* to 8 *kOe*. Following this field-sweeping sequence, $B_S^+$ ($B_S^-$) and $B_F^+$ ($B_F^-$) in the *MR(B)* curve define the transition fields from helical to the skyrmion, and the skyrmion to ferromagnetic phase, respectively[24-26], or vice versa, where the superscript "+ (−)" denotes the positive (negative) branch of the sweeping field. In contrast to the continuous *MR(B)* curves for thick wires (above 200 *nm*)[17], this *MR(B)* curve displays two discontinuous jumps with almost the same amplitudes in the skyrmion phase of $B_S<B<B_F$. This difference comes from the fact that the skyrmion lattice cannot be supported in the cross-section of such a narrow *NW*. Instead, a skyrmion cluster state (*SC*) with sparsely distributed skyrmions is present[18,19]. As the skyrmion is a topologically stable spin texture with an apparent "particle" character, a single skyrmion in the *SC* cannot be created or destroyed by smooth variation of local moments from other phases. Thus a field-driven cascading quantized transition is a natural result when the number of skyrmions, $N_s$, changes *one by one* under *B*. The maximum number, $N_{max}$, in a skyrmion cluster state can be easily estimated[18,19]. As discussed above, the presence of a merohedral twin boundary splits the parallelogram cross-section into two triangles[20]. Simple geometry analysis suggests one skyrmion at most can exist in each triangle, resulting in a total number of $N_{max} = 2$. This number perfectly matches the number of discontinuous jumps (or drops) in *MR*.

This interpretation of the *MR* results is further supported by Monte Carlo simulations based on the actual sample geometry, in which the 40 *nm NW* is divided into two merohedral lattices. Details of the model are described in the numerical methods section of the Supplementary Information. Snapshots of both three-dimensional and the cross-sectional spin configurations are, respectively, shown in the upper and bottom part in Figure 3a-3e, where the magnetic field *B* has the unit $B_S=J/\mu_B$, with *J* the exchange constant and $\mu_B$ the Bohr magneton. At a low field, a distorted helical order is established with the propagation direction lying in the cross-section and perpendicular to the twin boundary (Figure 3a). This is different from the conventional bulk sample, where the propagation is along the <111> direction due to weak crystal anisotropy[1]. This difference comes from the presence of the twin boundary, at which the *DM* interaction vanishes and only ferromagnetic exchange interaction survives. Therefore ferromagnetic ordering is persistent along the twin boundary plane, and modulation along the *NW* is prohibited. When the field $B/B_S$ is increased above a threshold, the skyrmion cluster appears (Figure 3c). Due to the geometric confinement, only one skyrmion is allowed in each merohedral domain and the skyrmions are aligned along the *NW*, forming two skyrmion tubes. At the *NW* boundary, spins align parallel to the boundary due to the *DM* interaction and missing spins near the boundary[18,19,27]. Consequently, the swirling direction at the boundary of each domain is opposite to that of the skyrmion within, and each skyrmion tries to sit at the center of each domain owing to the repulsion from the edge states[6,7]. Notice that before entering skyrmion cluster state, the system undergoes an intermediated state (Figure 3b), characterized by forming half-skyrmions in the interior of the nanowire. As a result, this state weakens the abrupt transition from the helical state into the skyrmion phase, being consistent with the unapparent transition field $B_S$ in Figure 2. With further increase of the field strength, one skyrmion disappears, leaving a mixed state of the 3D modulation in one domain and a skyrmion tube in the other domain (Figure 3d ). Although in the simulations, the geometries of the two domains in the nanowire are identical except for different chiralities, the ferromagnetic exchange at the twin boundary gives rise to certain correlations between two domains. Our result shows that the formation of 3D modulations in one domain can stabilize the remaining skyrmion in the other, so that two skyrmions do not disappear simultaneously. In real samples, the sizes of the two twins are not the exactly identical, which may also lead to the difference in the corresponding critical fields. At an even larger field, the remaining skyrmion transforms into the fully polarized ferromagnetic state (Figure 3f) via distorted conical phase in a tiny magnetic field interval (Figure 3e), while swirling along the boundary persists (Figure 3f). The phase diagram for these states is shown in Figure 3g. This two-step destruction of the *SC* state corresponds to the two jumps observed in the *MR(B)* curve.

The resistance jumps are persistent at various temperatures, as shown in Figure 4a. However they gradually evolve into kinks at higher temperatures, because thermal fluctuations smear out the topological transitions. Although the kinks become indistinct above 19 *K*, a closer examining of the *dMR/dB* data (Figure 4b) still shows two peaks between $B_S$ and $B_F$, indicating the survival of the two-skyrmion cluster. Symmetric behaviors are shown in both positive and negative branches of the sweeping fields at high temperatures. However, when the temperature drops below 15 *K*, the magnetization kinetics plays an increasingly important role, as indicated by the reappearance of the conical phase, shown by kinks instead of jumps observed at $B_F$ in the negative field branches[17]. This conical phase survives in a narrow window sandwiched between $B_F$ and $B_C$ shown in Figure 4a.

Figure 4c shows the *T-B* phase diagram constructed from transport measurements, in which the solid symbols are derived from $B_S$, $B_c$, and $B_F$ in the *MR(B)* curves and the open symbols come from the transition temperature $T_{S1}$ and $T_{S2}$ in the *MR(T)* curves at fixed *B* (Supplementary Figure S3). It is found that skyrmion clusters with different $N_s$ are stabilized widely in the *T-B* plane, ranging from the $T_c$ down to 5 *K* on the positive field branch, while eclipsed by the conical state in a narrow window at low temperatures on the negative branch. The saturation field $B_F$ is lower than that previously reported (*6 kOe* at low temperatures) for bulk MnSi crystals[21] and thick MnSi wires[17].

To further understand the importance of *NW* size and geometric confinement, we systematically investigated the transport properties of *NWs* with different diameters. Figure 5a-5c show typical *MR* curves for three *NWs* with diameters of 80 *nm*, 55 *nm* and 20 *nm* at representative temperatures. The corresponding *dMR(B)/dB* data are simultaneously plotted as gray dotted lines to identify the phase transitions clearly. More discontinuous jumps are observed at a thicker *NW* with a diameter of 55 *nm*. A detailed analysis clearly shows three additional jumps with similar jump height in the interval $B_S < B < B_F$, as labeled in Figure 5b. A similar numerical calculation shows that only two skyrmions can exist in each triangle divided by the twin boundary, and therefore $N_{max}=4$, which perfectly matches the number of jumps (see Figure S7). Comparing Figure 5b and Figure 4a, we see that the thinner the *NW* is, the more prominent the discontinuous jumps are. This can be understood phenomenologically from the anisotropic magnetoresistance (*AMR*)[28], which obeys $\rho(\theta) = \rho_0 - \rho_A \cos^2\theta$, where $\theta$ is the angle between local moments and the cross section, and the residual resistance $\rho_0$ is much larger than the anisotropic *MR* $\rho_A$. The presence of a skyrmion leads to the spatial modulation of local moments, average over which gives the *MR* of the wire. It can be easily shown that the resistance change once we place a single skyrmion is given by $\delta\rho = c(R^2/d^2)(\rho_A/\rho_0)$, where $R$ is the skyrmion radius, and $c$ is a dimensionless quantity of unity order. Therefore the larger the sample size is, the smaller the jump height will be, as shown in Figure 5b. We also notice that the jumps in the negative branches are more obvious than those in the positive ones, reinforcing the important role that the magnetization kinetics plays. The observed exceptional data points in the jump region may originate from the emergence of singularities during the topological transition, such as monopoles suggested by the magnetic force microscopy (*MFM*) image of bulk $Fe_{0.5}Co_{0.5}Si$[29]. In addition, several transitions in the interval $(B_S^+, B_S^-)$ are also observed. The simulation results suggest these phases are highly relevant to emergence of the elongated skrymion, i.e., bimron, The image of which is shown in Figure S7 of the supplementary information.

For an even thicker *NW* with a width of 80 *nm*, the isothermal *MR* curves display completely different transport behavior from the narrow *NWs*, but corroborate the well-established scenario for helimagnets[4,17]: the helical phase (H) is stable at low fields ($B < B_S$), and then transits into the skyrmion state (S) at $B_S$ via a discontinuous phase transition. At even higher fields ($B > B_C$) the conical phase appears, and eventually turns into the ferromagnetic ordering above $B_F$. As the *NW* diameter ($d \sim 80$ *nm*) is much larger than the single skyrmion size ($L_d \sim 18$ *nm*), closely packed skyrmions are hosted, leading to a skyrmion lattice, rather than a skyrmion cluster, and the continuous *MR(B)* curve between $B_S$ and $B_F$. This is in contrast to the jumps in *MR* seen in narrow *NWs* due to the emergence of skyrmion clusters.

As a proof, in the other limit when the *NW* is extremely narrow, the discontinuous jumps indicating the number of skyrmions should be absent. To this end, a *NW* with a width of ~ 20 *nm* is examined. The *MR* follows a smooth curve with two transition fields (marked as $B_S$ and $B_C$) in the whole magnetic field region (Figure 5c). This is expected for such narrow *NW*s because each small triangular fails to host a single skymion (with a size of ~ 18 *nm*). These experimental observations are also reproduced by the Monte Carlo simulation, as shown in Figure S8, where only helical, distorted conical and field-polarized ferromagnetic states exist. However, the calculated transition fields are much smaller than these inferred from the experimental measurements. On the other hand, compared with the thicker nanowires, we observed the high critical temperature and magnetic hysteresis behavior around $B_S$ at low temperature. Previous experimental work identified that the critical ordering temperature is strongly affected by finite size effect in *NW*s[16]. A similar hysteresis was also observed in the easy-plane B20 MnSi/Si(111) film when the magnetic field is applied along the plane[30]. These similarities suggest the important role of finite size effects for such ultra-narrow wires. It has been extensively discussed that the change of lattice structure at the crystal boundary or surface contributes an additional an additional magnetic anisotropy or interfacial *DM* interaction[31]. Because of the large surface-to-volume ratio in the ultra-narrow nanawire, we expect these additional interactions play a significant role in determining the magnetic phases and magnetization process. Nevertheless, based on the *MR* data and our simple parameter-free model, it is not possible to conclusively decide which magnetic structures are responsible for the field-driven transitions in the *MR* data. Therefore, based on the numerical results, we only suggest possible helical and distorted conical phases separated by $B_S$. This is also supported by the fact that the *MR* curve appears to be similar to that observed in bulk MnSi at low temperatures[25], where only helical, conical and field-polarized ferromagnetic states exist. Discussions about the additional surface interactions are beyond the scope of this work.

On the basis of the *MR* data for these *NW*s at temperatures below $T_c$ (see the detailed data in Figure S4), we can summarize the *T-B* phase diagrams for *NW*s with different diameters, as shown in Figure 5d-f. For $d\sim 20$ *nm*, the critical field $B_F$ is much larger than that in bulk MnSi (Figure 5f). Importantly, as the size of the *NW* increases, a skyrmion cluster phase emerges. At $d \sim 40$ *nm*, which is about two times larger than $L_d$, for example, the skyrmion cluster state is dominant while the conical state appears in a tiny *T-B* window on the negative branches at low temperatures (Figure 4c). As the *NW* diameter further increases, individual skyrmions are closely packed into a lattice (Figure 5d) through a crossover region (Figure 5e).

In summary, we have demonstrated the presence of skyrmion cluster states in confined MnSi nanowires with diameters comparable to single skyrmion domain size. The maximum number of skyrmions within this cluster is determined by the dimension of the nanowire cross sections, and the skymion number can be controlled by external magnetic fields and revealed by quantized jumps in the *MR* curves. These results not only reveal new physics of the skyrmion cluster states in confined geometries, but also can guide the development of skyrmion-based memory devices in which the individual skyrmions could be utilized for multibit memory cells.

**Methods Summary**

**Device fabrication**

All MnSi nanowires (*NW*s) used in this work were synthesized by chemical vapor deposition using $MnCl_2$ precursor onto silicon substrates[14]. To fabricate *NW* devices for magnetotransport measurements, a single *NW* was picked up from the original silicon substrates and transferred onto a new clean Si substrate coated with 300 *nm* silicon oxide by three-axis hanging joystick oil hydraulic micromanipulators under an optical microscope, where a home-made tip was used to pick up the *NW* (See Figures S1 in Supplementary Information). Four Pt or W electrodes were patterned onto an individual *NW* using focused-ion beam (FIB) techniques with a small beam current of 7.7 *pA* or 24 *pA,* using a FEI Helios Nanolab 600i. Prior to Pt deposition, the contact areas were milled by about 5 *nm* with a beam current of 1.1 *pA* in order to remove surface oxide. To avoid contamination of the electrode and destruction of the sample by the Ga ion source, a 200 *nm* thick PMMA resist was used to cover the rest of the samples between the electrodes by regular e-beam lithography technique prior to the fabrication of the electrodes (See Figure S2 in Supplementary Information).

**Measurement methods**

Standard 4-probe transport measurements on individual *NW*s were carried out in a Physical Property Measurement System (PPMS, Quantum Design Inc.) with a 16 T superconducting magnet. For the transport measurements, to obtain good data with high signal-to-noise ratio, large currents of 30 $\mu A$, 10 $\mu A$, 5 $\mu A$ and 2 $\mu A$ were used to measure the 80 *nm*, 55 *nm*, 40 *nm* and 20 *nm* *NW*s, respectively. Transport data under small currents were also measured to make sure that the effects of joule heating on the experimental results due to large current density were negligible. All *MR(B)* data were recorded by increasing *B* from -8 *kOe* to 8 *kOe*. The sweeping fields corresponded to the properties of *NW*s under field cooling (*FC*). We also recorded the initial *MR* curves, reflecting the properties of *NW*s under zero field cooling (*ZFC*) (by cooling the *NW*s to the defined temperature from far above $T_c$ , e.g., ~ 50 *K*, in zero magnetic field before recording the *MR* data). However, this difference between *ZFC* and *FC* does not significantly affect the final results (See Figure S5 in Supplementary Information).

**Acknowledgments**

This work was supported by the National Key Basic Research of China, under Grant No. 2011CBA00111; the National Nature Science Foundation of China, Grant No. 11174294, No.11474290, No. 11104281, No. 11374302, No. 11222434, No. U1432251; the Anhui Provincial Nature Science Foundation of China, Grant No.1308085QA16; National Basic Research Program of China, Grant No. 2013CB934600; US National Science Foundation, Grant No. ECCS-1231916, No. ECCS-1408168; US Department of Energy, Grant No. DEFG02-08ER46544; and the Theoretical Interdisciplinary Physics and Astrophysics Center. The authors thank Weike Wang and Yuyan Han for assistance with the sample measurements by PPMS.

**Author contributions**

S. J. and M. T. supervised the research. M. J. S. and D. L. synthesized the MnSi nanowires. H. D. and C. J. fabricated the devices. H. D., C. J., J. Y., Y. X., W. N., and J. W. carried out the transport measurements in PPMS. L. K. and J. Z. performed the calculations, and J. Z. wrote a significant part of the discussion. R. C. performed TEM analysis. H. D., M. T., J. Z., D. L. Y. Z. and S. J. contributed to the analysis. H. D., M. T., J. Z., D. L., M. J. S. and S. J. wrote the paper, together with help of all other co-authors.

multilayers . *Phys. Rev. Lett.*, **87**, 037203 (2001)

**Figure Captions:**

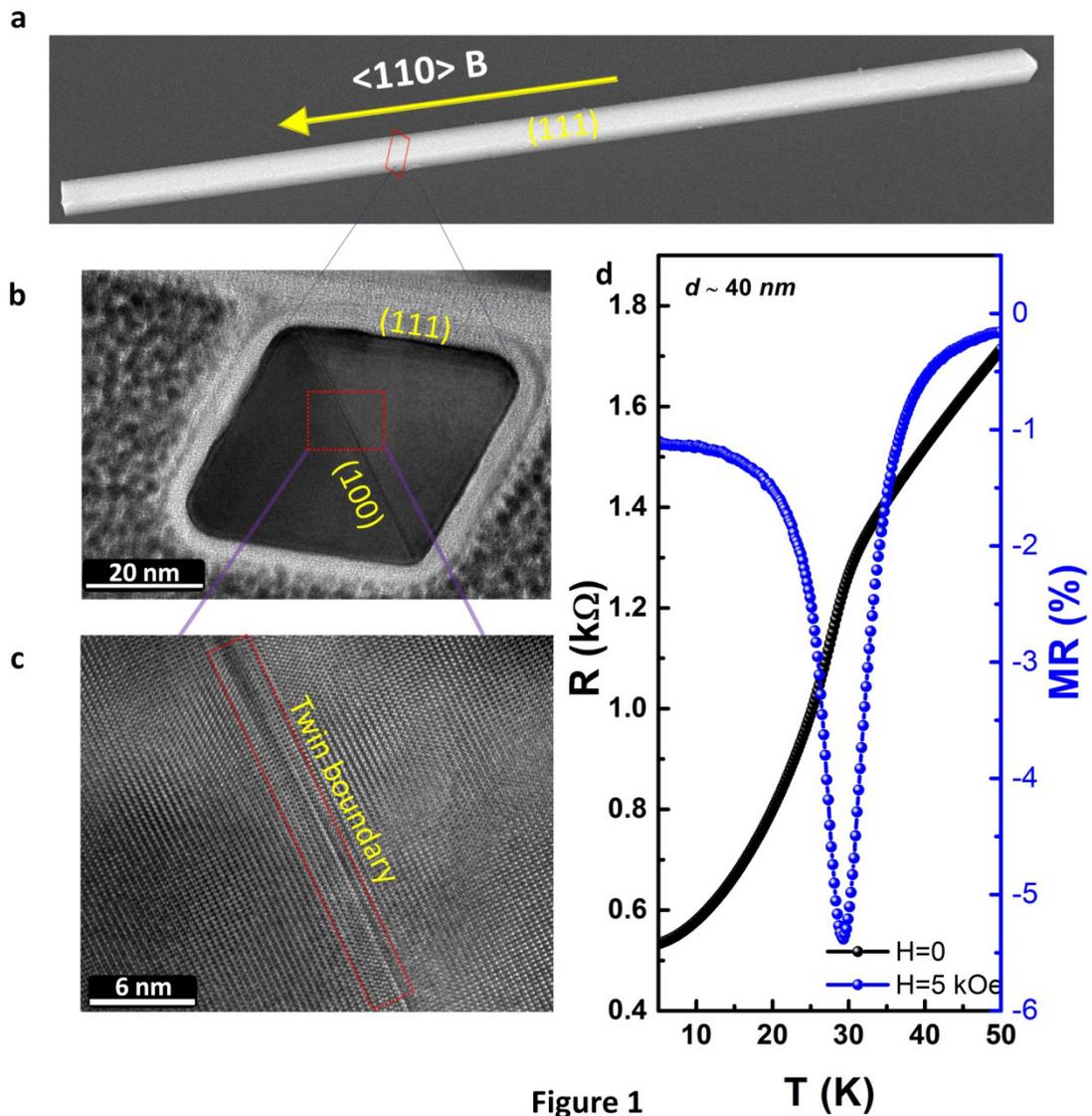

Figure 1

**Figure 1 Crystal morphology and transport properties of MnSi nanowires. a**, A typical SEM image of a MnSi *NW* with a smooth (111) surface and <110> growth direction. **b**, A cross-sectional TEM image of a *NW* with a merohedral twin boundary, where the (001) twin plane is parallel with the <110> growth direction. **c**, High resolution TEM image of the cross-section at the twin boundary. **d**, Temperature dependence of the resistance (black*)* and magnetoresistance (*MR*) (colored dots) at 0 and 5 *k*Oe, showing a $T_c$ of 29 *K*.

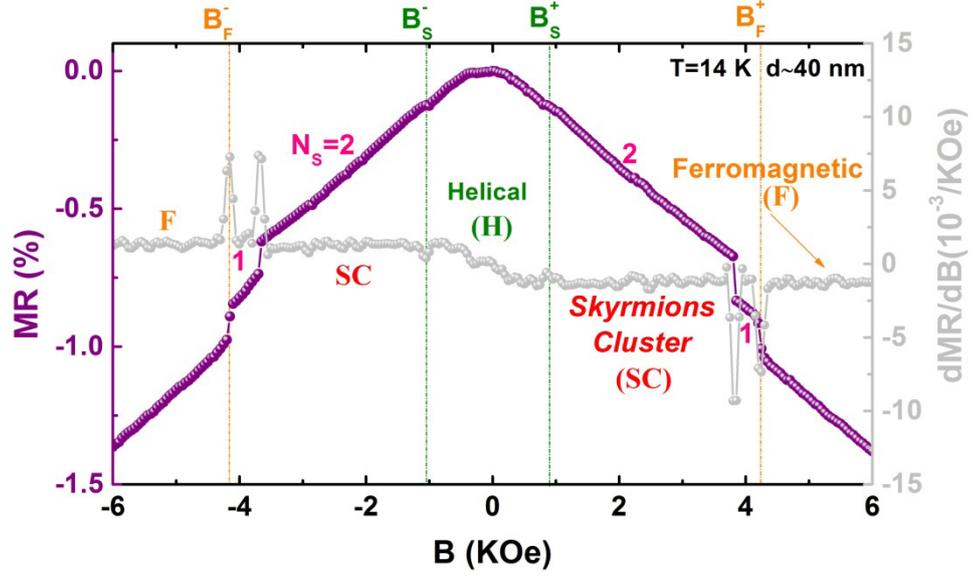

**Figure 2**

**Figure 2 Variation of spin configurations with magnetic fields for a 40 *nm* NW.** *MR* (colored dots) and *dMR/dB* (light gray dots) versus the magnetic field *B*. The measurements were carried out by increasing the field from -8 *kOe* to 8*kOe*. $N_S$ represents the number of skrymions in the skyrmion cluster states. The transition fields, extracted from the *dMR/dB* data, at positive and negative sweeping magnetic field branches are denoted by the superscripts "+" and "−", respectively. H, SC, C and F stand for the helical, skyrmion clusters, conical and ferromagnetic phases, respectively.

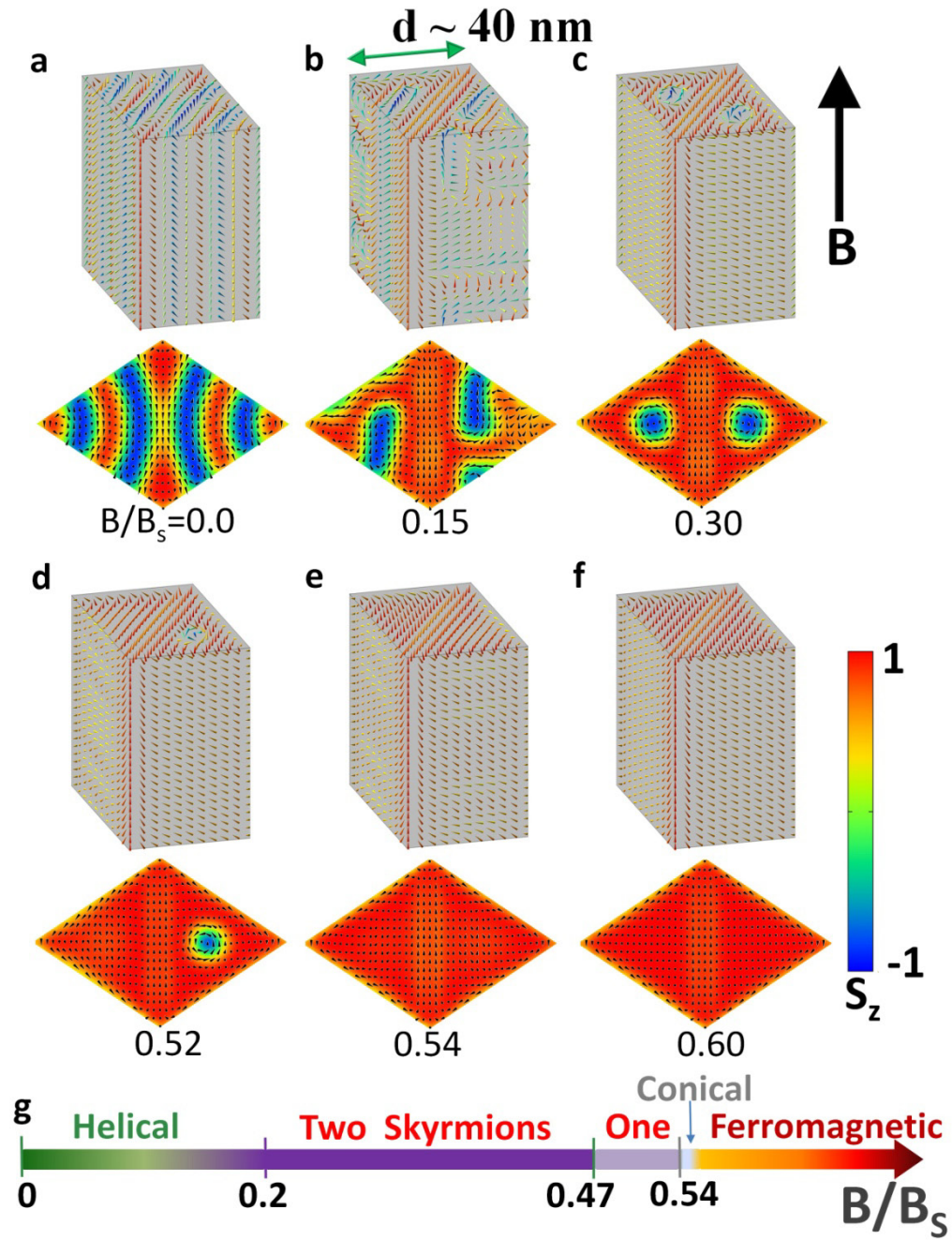

**Figure 3 Calculated spin configurations for the typical states in the 40 *nm* wire. a**: distorted helical; **b**: intermediated state with meron in the interior of the *NW*; **c**: two skyrmions; **d**: one skyrmion; **e**: distorted conical phase or 3D modulations; **f**: field-polarized ferromagnetic state ) and **f-i,** the corresponding cross-sections of these states. **g**, The phase diagram in *B* space, where $B_S=J/\mu_B$ is the unit of magnetic field in the Monte Carlo simulations.

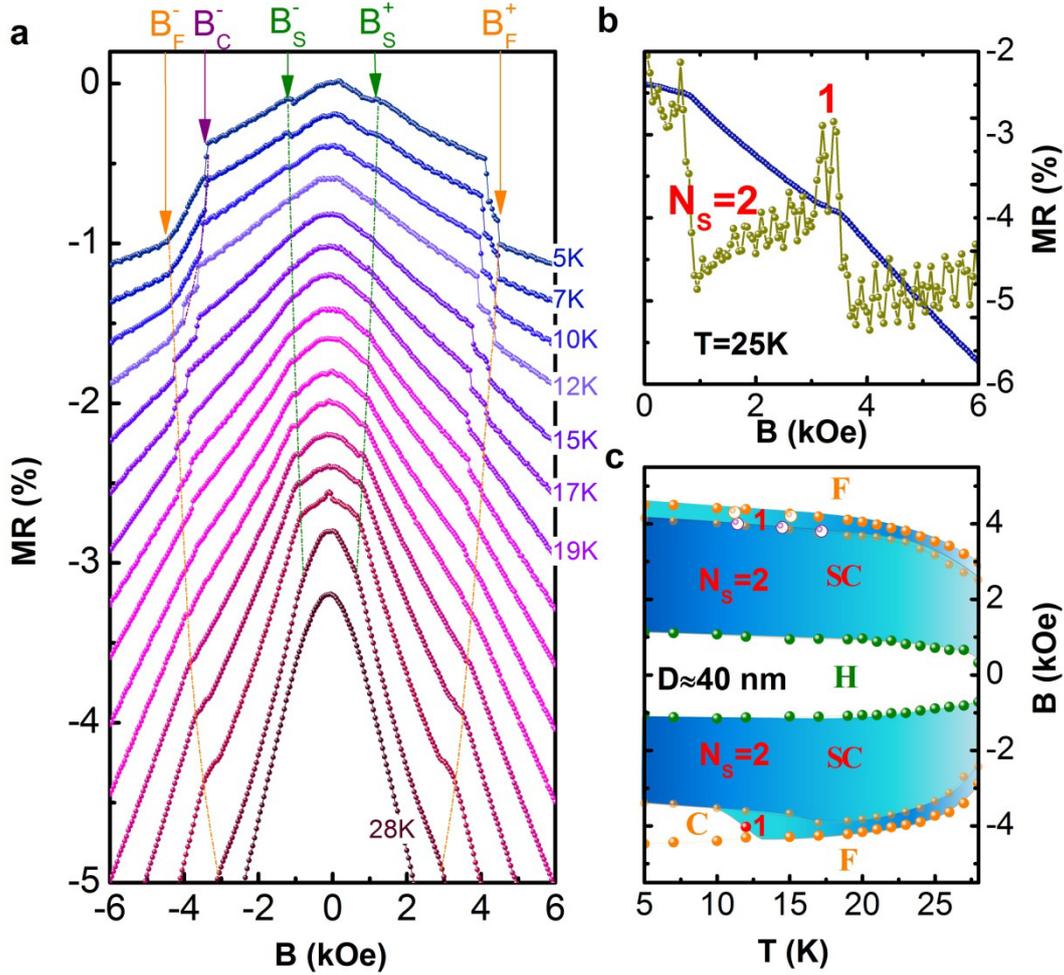

**Figure 4**

**Figure 4 Transport properties and *T-B* phase diagram of a 40 *nm* MnSi *NW*. a,** *MR* at various temperatures below $T_c$. Three transition fields are marked by different colors. The data are shifted for clarity and the temperature is increased in 1 *K* steps from 19 to 28 *K*. **b**, *MR* and *dMR/dB* as a function of *B* at a high temperature *T* = 25 *K*, where two peaks in the *dMR/dB* indicate the persistence of two individual skyrmions. **c,** Phase diagram inferred from *MR* data. Solid and open symbols represent data obtained from *MR(B)* and *MR(T),* respectively. $N_S$ stands for the number of skyrmions in the cluster. Individual skyrmions survive in a large *T-B* window down to the lowest measured temperature (5 *K*) in the positive magnetic branches.

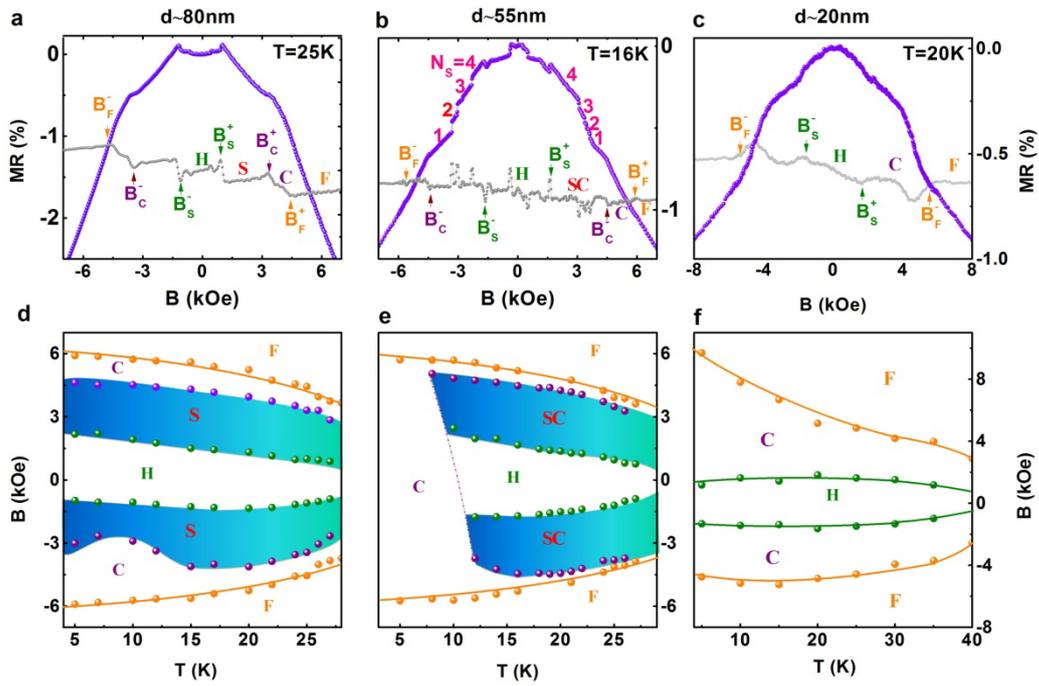

**Figure 5**

**Figure 5 The *NW* diameter dependence of the MR behaviors and the skyrmion phase diagrams of temperature *T* and magnetic field *B*. a-c,** *MR* as a function of *B* for three *NW*s with different diameters (colored dots; **a**, 80 *nm*; **b**, 55 *nm*; **c**, 20 *nm*). The transition fields are extracted from the corresponding *dMR/dB* data (light gray dots). **d-e**, The phase diagrams obtained from the *MR(B)* data. S stands for the magnetic state with lots of number of skyrmions. The colored regions indicate the skyrmion states.

# Supplementary Information
# Electrical Probing of Field-driven Cascading Quantized Transitions of Skyrmion Cluster States in MnSi Nanowires

*Haifeng Du,[1] Dong Liang,[2] Chiming Jin,[1] Lingyao Kong,[3] Matthew J. Stolt,[2] Wei Ning,[1] Jiyong Yang,[1] Ying Xing,[4] Jian Wang,[4] Renchao Che,[5] Jiadong Zang,[6]\* Song Jin,[2]\* Yuheng Zhang,[1,7] Mingliang Tian[1,7]\**

[1]High Magnetic Field Laboratory, Chinese Academy of Science (CAS), Hefei, Anhui Province 230031, China

[2]Department of Chemistry, University of Wisconsin—Madison, 1101 University Avenue, Wisconsin 53706, USA

[3]Institute of Fluid Physics, China Academy of Engineering Physics, Mianyang, Sichuan Province 621900, China

[4]International Center for Quantum Materials, School of Physics, Peking University, Beijing 100871, China

[5] Advanced Materials Laboratory, Fudan University, Shanghai 200433, P. R. China

[6]Institute for Quantum Matter and Department of Physics and Astronomy, Johns Hopkins University, Baltimore, Maryland 21218, USA

[7] Collaborative Innovation Center of Advanced Microstructures, Nanjing, Jiangsu Province 210093, China

*Corresponding authors: jiadongzang@gmail.com (J.Z.); jin@chem.wisc.edu (S.J.); tianml@hmfl.ac.cn (M.T.).

## I. Home-made microtip for the transfer of nanowires

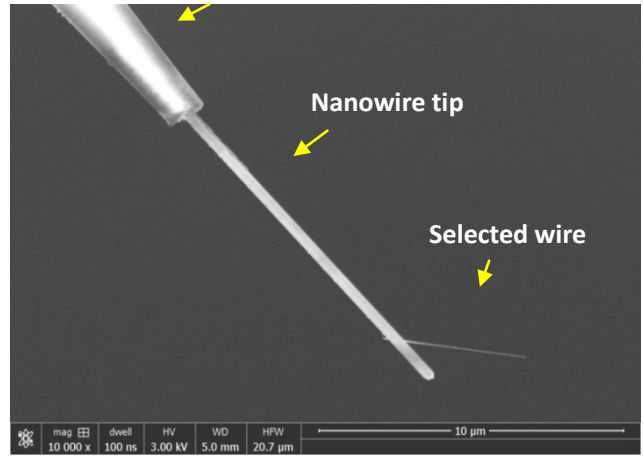

**Figure S1. Microtip for the transfer of nanowires.** A thick nanowire is inserted into a hollow glass tube. This wire is used as a new tip instead of the capillary glass tube for picking up the nanowire samples. In this way, the success rate for selecting the desired nanowire is significantly improved. A thin nanowire is stuck on the thick nanowire at its tip.

## II. Standard four-probe nanowire device for transport measurements

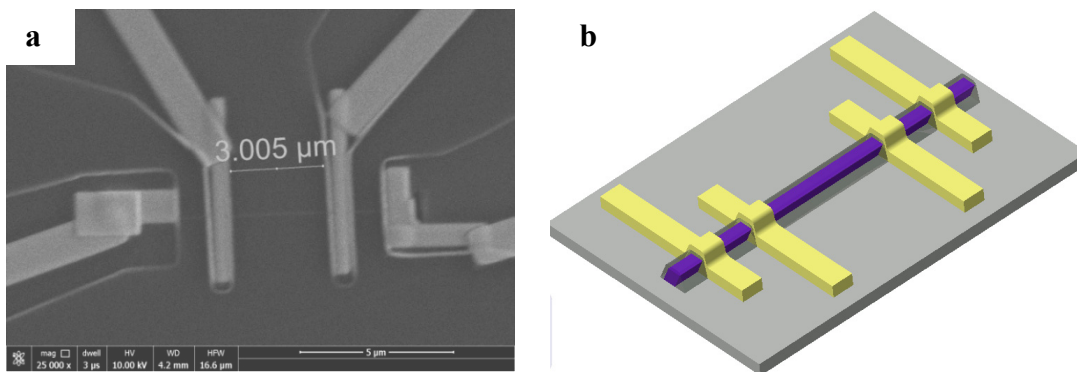

**Figure S2. Four-probe electrical device for transport measurements. a,** A typical scanning electron microscopy (SEM) image of a MnSi nanowire device. We first used e-beam lithographic technique to expose the PMMA resist in the intended areas of the electrodes. After the developing process, we got the well-defined e-beam pattern on the nanowire. In order to achieve good ohmic contact with negligible contact resistance to the MnSi nanowires, we deposited four Pt electrodes by focused

ion beam (FIB) technique at the areas defined by e-beam, the rest of the nanowire samples between the electrodes are still covered by 200 *nm* thick PMMA e-beam resist for protection of the nanowire. **b**, A schematic illustration of the device with four electrodes (yellow stripe) and MnSi wire (purple).

**III. Magnetoresistance (*MR*) versus temperature (*T*) curves of the 40 *nm* nanowire at different *B***

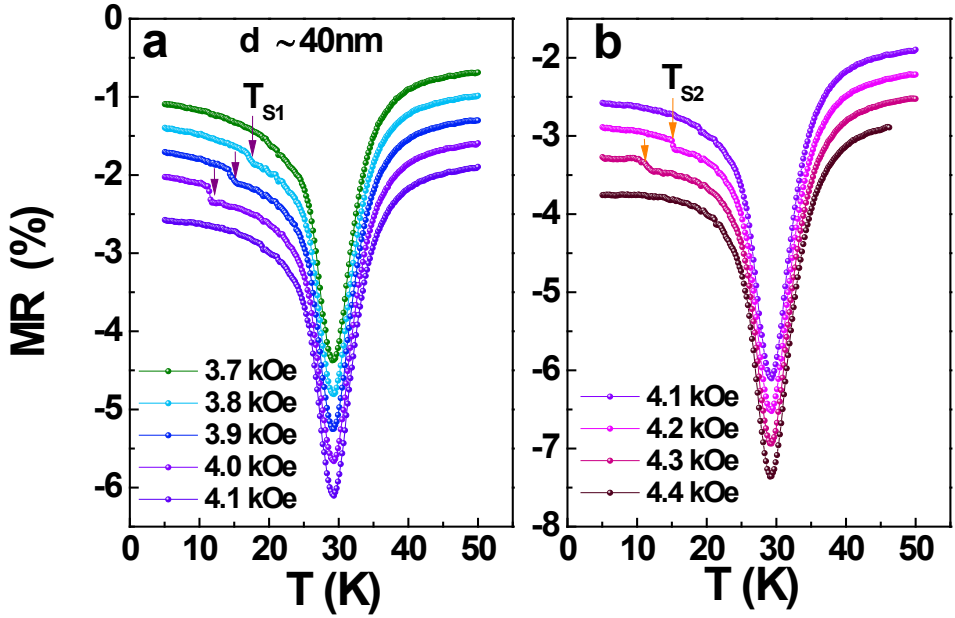

**Figure S3. Temperature dependence of the *MR* for the 40 *nm* nanowire under different magnetic fields. a,** The magnetic fields from 3.7 *kOe* to 4.1 *kOe*; **b,** The magnetic fields from 4.1 *kOe* to 4.4 *kOe*. A deep valley around $T_c$ is seen due to the suppression of carrier scattering from spin fluctuations. Below $T_c$, a more complex *MR* behavior that sensitively varies with the increase of *B* is seen between 3.7 *kOe* and 4.4 *kOe*. A comparative analysis of the *MR* vs. *T* and *MR* vs. *B* data allows us to conclude that $T_{S1}$ corresponds to the lower critical temperature from two skyrmions into one, while $T_{S2}$ corresponds to the transition temperature from one skyrmion into the ferromagnetic state. All data are shifted for clarity.

**IV. Magnetoresistance versus magnetic field data for 80 *nm*, 55 *nm* and 20 *nm* nanowires at different temperatures**

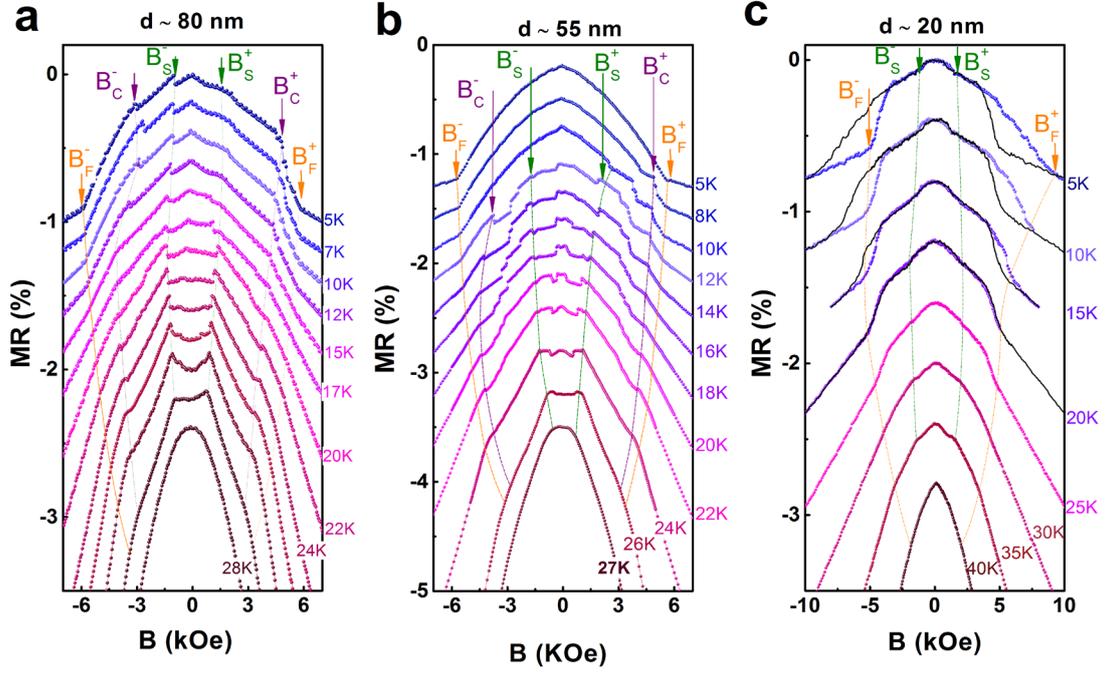

**Figure S4.** *MR* as a function of *B* for three nanowires with different diameters (a, 80 *nm*; b, 55 *nm*; c, 20 *nm*) below $T_c$. The transition fields $B_S$, $B_C$ and $B_F$ are defined the same as in the main text. The dotted lines mark the phase transitions for the guiding the eyes. For the 80 *nm* nanowire, *MR* curves show continuous change in the skyrmion state between $B_S$ and $B_C$, while discontinuous ones are observed for the 55 *nm* nanowire. Further decreasing the size of the nanowire to 20 *nm* caused the skyrmion state to disappear. All data are shifted for clarity. The black lines in **c** show the data collected when the field was swept from positive to negative and significant hysteresis loops appear below about ~ 20 *K*.

## V. Magnetoresistance versus magnetic field curves for the 40 *nm* nanowire at different temperatures under zero field cooling (*ZFC*)

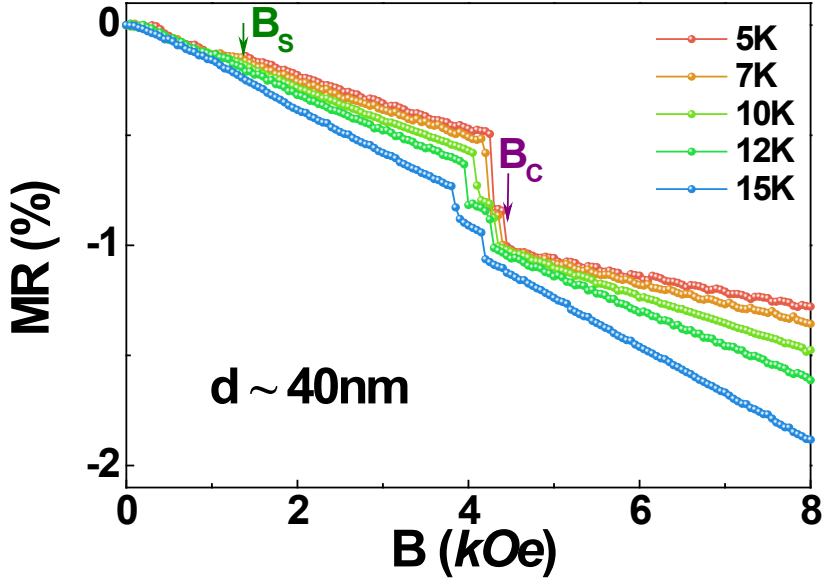

**Figure S5. The magnetic field dependence of the *MR* for the 40 *nm* nanowire under *ZFC*.** To obtain the data under *ZFC*, we raised the temperature far above $T_c$ (e.g., ~ 50 *K*) in zero magnetic field for each measurement, and then cooled the nanowires to the defined temperature in zero magnetic field before recording the *MR* data from zero field to high fields. It is noted that above 15 K, the curves under *FC* and *ZFC* are the same. Below 10 K, the *MR* curves after *ZFC* show a close agreement with that in the positive branch under *FC*.

## VI. Numerical method: Monte Carlo simulations

In order to support the experimental observations, Monte-Carlo (*MC*) simulations were performed to get a clear picture of the spin configurations. A generally accepted Hamiltonian for chiral magnets including direct and indirect magnetic coupling with constants $A$ and $D$, magnetization $\bm{m}$, and external field energy is written as[S1]

$$w = A(\nabla \bm{m})^2 + D\bm{m} \cdot (\nabla \times \bm{m}) - \bm{m} \cdot B \qquad (1)$$

For the *MC* simulation, the 40 *nm* nanowire is divided into two discrete blocks with unit magnetization $\hat{S}$, then, a three dimensional (*3D*) lattice Hamiltionian corresponding to Eq.(1) is written as

$$E = \sum_{<ij>}(-J\hat{S}_i \cdot \hat{S}_j + D\hat{r}_{ij} \cdot (S_i \times S_j)) - \sum_i S_i \cdot B^r + E_c \qquad (2)$$

Where $J$ denotes the ferromagnetic exchange coupling constant; $D_{R_{ij}}$ is $DM$ interaction constants with $|D_{R_{ij}}| = D$ and $D_{R_{ij}}$ is the vector pointing along site $i$ and $j$; $<ij>$ denotes the nearest spins. $B^r$ stands for the reduced magnetic field. The constructed model is schematically illustrated in Figure S6.

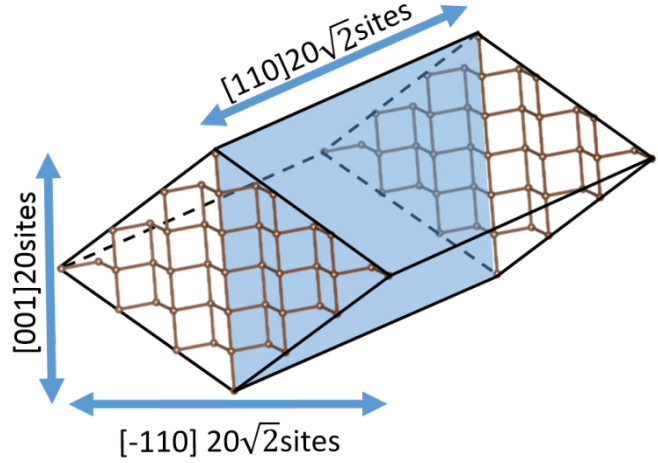

**Figure S6. The orientation of the simple cubic lattice used in our MC simulations.** The blue shaded areas mark the twin plane.

In the discrete model both direct and $DM$ exchange in an inhomogeneous state will have slightly different energy depending on their orientation with respect to the underlying discrete bond orientation of the crystal lattice. Contrary to the isotropic continuous model in Eq. (1), the discrete model in Eq. (2) used in the $MC$ simulations leads to the anisotropy energy. To smear out this effect, a corrected term $E_c$ is added as[S2]

$$E = \sum_{<<ij>>}(-\frac{J}{16}\hat{S}_i \cdot \hat{S}_j + \frac{D}{8}\hat{r}_{ij} \cdot (S_i \times S_j)) \qquad (3)$$

where $<<ij>>$ represents the next nearest spins. The ratio $D/J$ was chosen to be consistent with the value adopted in Ref. S2, where the wavelengths of $d=10$ lattice constants were used.

Concerning the real geometry, open and periodic boundary conditions were adopted in the cross-sectional plane of the nanowire and the long axis, respectively, and a boundary forming the same nanowire geometry as the experimental merohedral twin plane, which divides the nanowire into two equal parts with opposite chirality. Thus, the sign of the coefficient *D* is opposite in the two domains. According to the previous Loretnz TEM observations in the another B20 compounds FeGe, the inversion of the lattice chirality (handedness) of the B20 structure across certain grain boundaries would lead to the invariance of the sign of the spin-orbit interaction within FeGe. It was also observed that the spins around the boundary are almost parallel with the grain boundary. This observation implies the ferromagnetic interaction would play dominant role in the narrow transition region where the inversion of the lattice chirality (handedness) occurs. yielding the nearly parallel spin arrangements in the thin region of crystal boundaries[S4]. Thus, the value of *D* is set to zero in the boundaries[S3]. A high temperature annealing metropolis algorithm is used to obtain the equilibrium spin configurations. At each temperature the system is allowed to relax towards equilibrium for the first $10^5$ Monte Carlo steps and thermal averages are calculated over the subsequent $10^5$ steps.

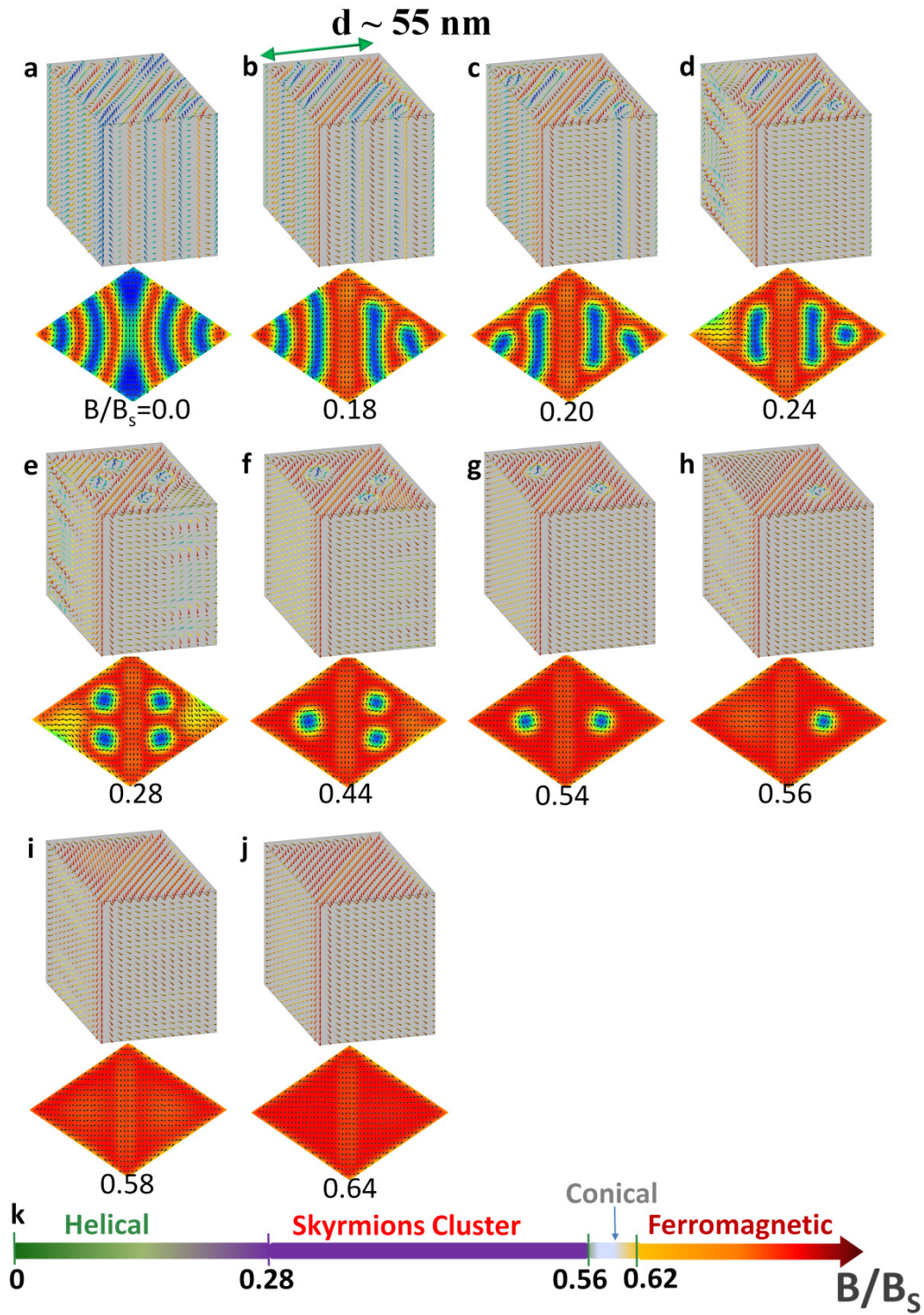

**Figure S7 Calculated spin configurations for the magnetic states in the 55 *nm* nanowire. a**: distorted helical state. **b-d**, intermediated state with the mixture of bimeron and skyrmions in the interior of the *NW*. The bimeron and skyrmion should respond to the observed jumps in the interval $(B_S^+, B_S^-)$ shown in Figure 5b. **e**, four skyrmions. **f**, three skyrmions. **g**, two skyrmions; **h**, one skyrmions. **i**, distorted

conical phase (3D modulation); **j**: field-polarized ferromagnetic state ). **k**, The phase diagram in *B* space.

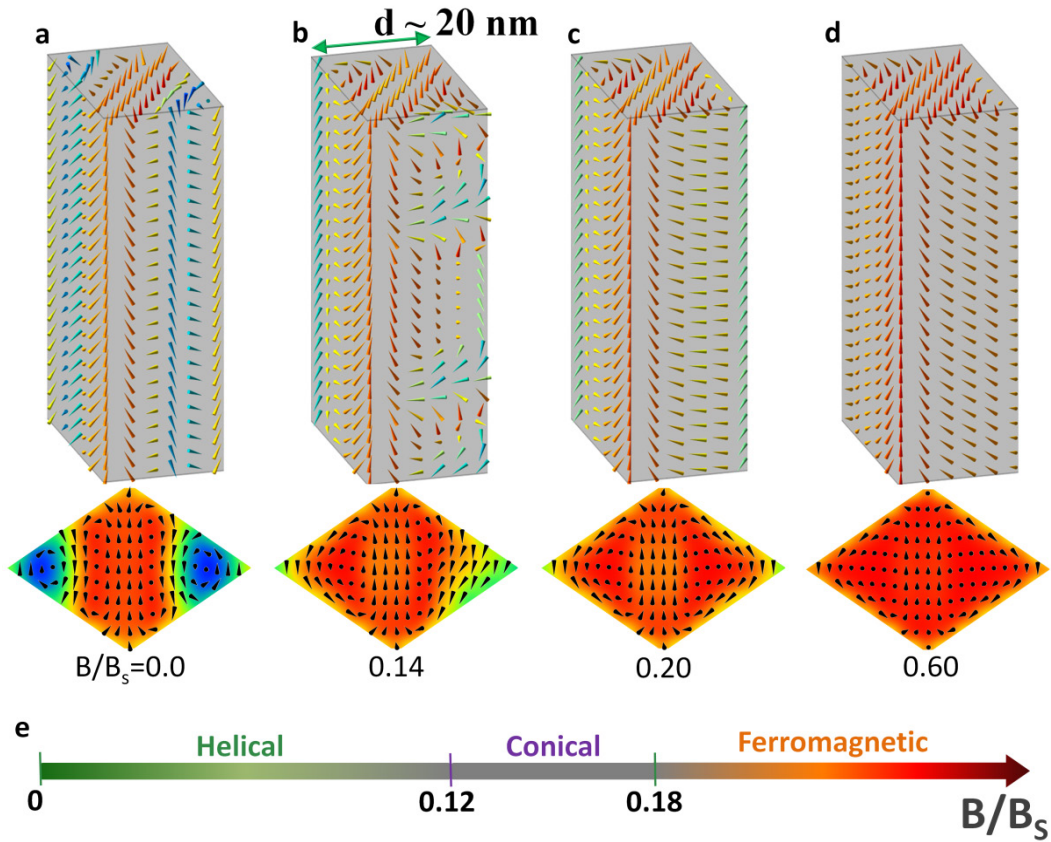

**Figure S8 Calculated spin configurations for the magnetic states in the 20** *nm* **nanowire. a**: distorted helical state. **b**, distorted conical phase (3D modulations); **c-d**: field-polarized ferromagnetic state ). **e**, The phase diagram in *B* space.